\documentclass{iopart}

\usepackage{setstack}
\usepackage{iopams}
\usepackage{acronym}
\usepackage{graphicx}
\usepackage{ifpdf}
\ifpdf
\else
\fi

\def\avtau#1{\left\langle #1 \right\rangle_\tau}
\def\vec#1{\mathbf{#1}}
\def\r{\vec{r}}

\def\x{\vec{x}}
\def\y{\vec{y}}
\def\ptsl{\xi_s}

\def\boltz#1#2{e^{-\beta H(#1,#2)}}
\def\boltzm#1#2{e^{-\beta \tilde H(#1,#2)}}
\def\Reg{{\cal R}}
\def\Zpart{{\cal Z}}

\acrodef{ABC}{amorphous boundary condition}
\acrodef{BC}{boundary condition}
\acrodef{CRR}{cooperatively rearranging region}
\acrodef{PBC}{periodic boundary condition}
\acrodef{RFOT}{random first-order theory}
\acrodef{SPM}{square plaquette model}

\begin{document}


\title{Numerical simulations of liquids with amorphous boundary
  conditions}

\author{A Cavagna$^1$, T S Grigera$^2$ and P Verrocchio$^3$}

\address{$^1$Istituto Sistemi Complessi (ISC-CNR), UOS Sapienza, Via
  dei Taurini 19, 00185 Roma, Italy, and Dipartimento di Fisica,
  Universit\`a\ ``Sapienza'', P.le Aldo Moro 2, 00185 Roma, Italy}

\address{$^2$Instituto de Investigaciones Fisicoqu{\'\i}micas
  Te{\'o}ricas y Aplicadas (INIFTA) and Departamento de F{\'\i}sica,
  Facultad de Ciencias Exactas, Universidad Nacional de La Plata,
  c.c. 16, suc. 4, 1900 La Plata, Argentina, and CONICET La Plata,
  Consejo Nacional de Investigaciones Cient{\'\i}ficas y T{\'e}cnicas,
  Argentina}

\address{$^3$Dipartimento di Fisica, Universit{\`a} di Trento, via
  Sommarive 14, 38050 Povo, Trento, Italy, and Istituto Sistemi
  Complessi (ISC-CNR), Via dei Taurini 19, 00185 Roma, Italy, and
  Instituto de Biocomputaci\'on y F\'{\i}sica de Sistemas Complejos
  (BIFI), Spain}

\date{August 31, 2010}

\begin{abstract}
  It has recently become clear that simulations under \acfp{ABC} can
  provide valuable information on the dynamics and thermodynamics of
  disordered systems with no obvious ordered parameter. In particular,
  they allow to detect a correlation length that is not measurable
  with standard correlation functions. Here we explain what exactly is
  meant by \acp{ABC}, discuss their relation with point-to-set
  correlations and briefly describe some recent results obtained with
  this technique.
\end{abstract}

\submitto{JSTAT}


\section{Introduction}

Systems that can be studied with numerical simulation are rather
small. Even those that can be considered ``very large'' by numerical
simulation standards are still quite small compared to macroscopic
experimental systems. Accordingly, a rather large fraction of the
particles is near the boundaries, with the consequence that \acp{BC}
have a significant effect on simulation results
\cite{book:rapaport04}. 
Very often, \acp{PBC} are used. These conditions avoid surface effects
(although of course finite-size effects are still present) and are
generally the most satisfactory way of studying bulk properties of
macroscopic systems \cite{book:rapaport04}.

Here we discuss a different kind of \acp{BC}, \emph{\acfp{ABC}}, which,
as has recently become clear, are highly useful in the study of
\emph{bulk} supercooled liquids \cite{mosaic:bouchaud04, self:prl07,
  self:nphys08} and likely other systems where it is hard to
detect spatial correlations due to lack of an obvious order parameter.

The basic idea, which is to investigate how far the effect of the
\acp{BC} penetrates into the bulk, has deep roots in statistical
mechanics. In fact, thermodynamic phase transitions can be
characterised as points where this penetration length is infinite, and
the \acp{BC} can be used to select one of several ergodic components
(coexisting phases) \cite{book:parisi98}. However this approach is not
normally used in simulations, as standard correlation functions or
fluctuations involving the order parameter are more convenient. In
systems like supercooled liquids, however, it is not clear how to
detect growing order. The novel idea \cite{mosaic:bouchaud04} is to
study the liquid under boundary conditions chosen by the system
itself (in a sense that will become clear in
sec.~\ref{sec:amorph-bound-cond}).

We discuss \acp{ABC} and discuss their close connection with
point-to-set correlations \cite{cs:mezard06,dynamics:montanari06} in
sec.~\ref{sec:amorph-bound-cond}. We then comment some recent
numerical results on the statics (sec.~\ref{sec:numer-results:-stat})
and dynamics (sec.~\ref{sec:numer-results:-dynam}) of liquids obtained with
\ac{ABC}. In sec.~\ref{sec:conclusions} we conclude.

\section{Amorphous boundary conditions and point-to-set correlations}

\label{sec:amorph-bound-cond}

It will be convenient to treat first the case of lattice systems (let
us call the lattice variables ``spins''), and to deal with the
continuum case afterwards. Consider an infinite system and chose a
region $\Reg$ of typical size $R$, which we shall call the cavity. Its
precise shape is not important, but it must be bounded; we shall
usually think of it as a sphere of radius $R$. We call $\sigma =
\{\sigma_1,\ldots,\sigma_M\}$ the spins belonging to the cavity
$\Reg$, and $\tau = \{\tau_{M+1},\ldots,\tau_N\}$ those outside. Let
$H(\sigma,\tau)$ be the hamiltonian of the infinite system.

To study the system under \acl{ABC} means to study the
(thermo)dynamics of the cavity $\Reg$ with the full hamiltonian
$H(\sigma,\tau)$ but with \emph{fixed} values of the outside spins
$\tau$, which act as a (disorderd) surface field. Naturally one has to
average all observables over different configurations of the outside
spins. Thus one computes what in disordered systems is usually called
the quenched average, which for the case of a one-time observable $A$
can be written
\begin{equation}
\overline{\avtau{A}} = \sum_\tau w(\tau) A_\tau,   \qquad 
 \avtau{A} = \frac{1}{\Zpart_\tau} \sum_{\sigma} A(\sigma,\tau)
  e^{-\beta H(\sigma,\tau)},
\end{equation}
with
\begin{equation}
 \Zpart_\tau =  \sum_{\sigma} e^{-\beta H(\sigma,\tau)},
\end{equation}
and we must specify the weight $w(\tau)$. The point is that this
weight is itself Boltzmann:
\begin{equation}
 w(\tau) = \frac{1}{\Zpart} \sum_\sigma e^{-\beta
  H(\sigma,\tau)} =\frac{\Zpart_\tau}{\Zpart},
 \qquad \Zpart =  \sum_{\sigma,\tau} e^{-\beta H(\sigma,\tau)}.
\end{equation}
In this sense the system ``chooses its own \acp{BC}''. The phrase is
even more appropriate when one considers the way
$\overline{\avtau{A}}$ is determined in a simulation: the system is
first equilibrated in a standard run, then an equilibrium
configuration is chosen and $A$ is measured in a simulation with the
spins outside the cavity artificially frozen (and the process is
repeated for several starting equilibrium configurations).

\subsection{One-time observables}

It is easy to see that \acp{ABC} do not affect the expectation value of
single-time observables. The annealed (Boltzmann)
average is
\begin{equation}
  \langle A \rangle = \frac{1}{\Zpart} \sum_{\sigma,\tau}
  A(\sigma,\tau) e^{-\beta H(\sigma,\tau)}.
\end{equation}
But writing the quenched average $\overline{A_\tau}$ in full we have
\begin{eqnarray}
  \overline{\avtau{A}} = \sum_\tau \frac{\Zpart_\tau}{\Zpart}
  \frac{1}{\Zpart_\tau} \sum_\sigma
  A(\sigma,\tau) e^{-\beta H(\sigma,\tau)} \nonumber\\
  = \frac{1}{\Zpart} \sum_{\tau,\sigma}  A(\sigma,\tau) e^{-\beta
    H(\sigma,\tau)} =
  \langle A \rangle .
  \label{eq:one-time}
\end{eqnarray}
Thus \emph{for one-time observables, the quenched and annealed
  averages are the same.} This includes in particular the
energy~\cite{personal:franz}, but also all non-connected $n$-point
static correlation functions, even those involving spins inside and
outside $\Reg$.

\subsection{Free energy, time correlations and overlap}

Quantities where \acp{ABC} make a difference are the free energy (as
one might expect) and, of more present interest, time correlation
functions. 

For the free energy one straightforwardly finds
\begin{eqnarray}
  \overline{F_\tau} &=& -\frac{1}{\beta} \sum_\tau w(\tau)
  \log\Zpart_\tau =  -\frac{1}{\beta} \sum_\tau w(\tau) \log\Zpart
  w(\tau) \nonumber\\
  &=& F -\frac{1}{\beta}\sum_\tau w(\tau)\log w(\tau).
\end{eqnarray}

Consider now a time correlation function. In the annealed case we have
\begin{eqnarray}
 C(t) &=&\langle A(t) B(0) \rangle \nonumber\\
  &=&\frac{1}{\Zpart} \sum_{\sigma,\tau\sigma',\tau'} A(\sigma,\tau)
  G(\sigma\tau|\sigma'\tau',t) 
  B(\sigma',\tau') \boltz{\sigma'}{\tau'},
  \label{eq:timecorr}
\end{eqnarray}
where $G(x|x',t)$ is the appropriate evolution operator, or
conditional probability. The corresponding fixed-$\tau$ expression is
\begin{equation}
  C_\tau(t) = \frac{1}{\Zpart_\tau} \sum_{\sigma\sigma'}
  A(\sigma,\tau) 
  G(\sigma\tau|\sigma'\tau,t) 
  B(\sigma',\tau) \boltz{\sigma'}{\tau},
\end{equation}
and the quenched average gives
\begin{eqnarray}
  \overline{C_\tau(t)} =\sum_\tau w(\tau) C_\tau(t)\nonumber\\
  = \frac{1}{\Zpart} \sum_{\sigma,\sigma',\tau}
  A(\sigma,\tau) 
  G(\sigma\tau|\sigma'\tau,t) 
  B(\sigma',\tau) \boltz{\sigma'}{\tau},
\end{eqnarray}
which differs from the annealed result~(\ref{eq:timecorr}) in that it
lacks the summation over $\tau'$.

In what follows we shall be particularly interested in the case
$A=B=\sigma_i$ (where $\sigma_i$ is a spin at the centre of $\Reg$)
and in the \emph{overlap}
\begin{equation}
   q \equiv \overline{\avtau{\sigma_i}\avtau{\sigma_i}} =
   \overline{\avtau{\sigma_i}^2}.
\end{equation}
The overlap is a measure of how much the \acp{ABC} restrict the
movement in phase space of the spin at the centre: the higher the $q$
(with respect to its annealed value) the higher the restrictions
imposed by the boundary. It is also the asymptotic value of the
spin-spin autocorrelation:
\begin{eqnarray}
  \overline{\big\langle \sigma_i(t)\sigma_i(0) \big\rangle_\tau} =
  \frac{1}{\Zpart} \sum_{\sigma,\sigma',\tau}
  \sigma_i \sigma'_iG(\sigma\tau|\sigma'\tau,t) 
  \boltz{\sigma'}{\tau} \nonumber\\
 \begin{array}{c}\longrightarrow\\\scriptstyle{t\to\infty}\end{array} \overline{\big\langle \sigma_i
    \big\rangle_\tau  \big\langle\sigma_i \big\rangle_\tau} =  q.
  \label{eq:autocorr-inf-time-limit}
\end{eqnarray}

\subsection{Point-to-set correlation}

The point-to-set correlation function between site $i$ and the
complement of $\Reg$ can be defined as \cite{dynamics:montanari06}
\begin{equation}
  C(i,\bar\Reg) = \langle \sigma_i f(\tau) \rangle, \qquad 
  \mbox{with}\quad  f(\tau)=\langle \sigma_i \rangle_\tau.
  \label{eq:point-to-set}
\end{equation}
Studying the value of $C(i,\bar\Reg)$ as a function of the size $R$ of
the cavity, one can define a characteristic length $\ptsl$ as the value
of $R$ beyond which the correlation drops below some small value. Part
of the interest of point-to-set correlations is that for models with
discrete variables on rather general graphs, rigorous bounds have been
proved showing that growth of correlation times must be accompanied
by a growth of $\ptsl$ \cite{dynamics:montanari06}. A divergence of
correlation times implies a divergence of $\ptsl$ even in models
where standard point-to-point correlations yield correlation lengths
that are always finite \cite{dynamics:montanari06,cs:mezard06}.
The correlation can be written
\begin{eqnarray}
  C(i,\bar\Reg) &=& \left\langle \sigma_i \avtau{\sigma_i} \right\rangle
   = \frac{1}{\Zpart} \sum_{\sigma,\tau} e^{-\beta
    H(\sigma,\tau)} \sigma_i
   \frac{1}{\Zpart_\tau} \sum_{\sigma'} e^{-\beta H(\sigma',\tau)}
   \sigma'_i \nonumber\\
   &=& \sum_\tau \frac{\Zpart_\tau}{\Zpart} 
   \left[ \frac{1}{\Zpart_\tau} \sum_{\sigma} \sigma_i
     \boltz{\sigma}{\tau} \right]
   \left[\frac{1}{\Zpart_\tau} \sum_{\sigma'} \sigma'_i
   \boltz{\sigma'}{\tau}\right] \nonumber\\
   &=& \overline{ \avtau{\sigma_i}^2} = q.
   \label{eq:point-to-set-expanded}
\end{eqnarray}
In this form it is clear that the point-to-set correlation is equal to
the overlap measured at the centre of the cavity. It is also clear
that it is related to sample-to-sample fluctuations of the spin
subject to \acp{ABC}, and it has the same form as the self-overlap
defined for spin glasses.

\subsection{The off-lattice case}

To rewrite the formulae of the previous section for the case of
off-lattice particles, one must introduce a hard wall enclosing the
cavity~\cite{personal:parisi}. The idea is that to write something
like the point-to-set correlation~(\ref{eq:point-to-set-expanded}) the
``set'' cannot be a volume to be filled with particles; it must be an
actual set of particles. In other words, one must choose the set based
on the \emph{identity} of the particles, and not on the values the
particles' degrees of freedom (choosing the particles that happen to
be within some particular volume is akin to building a set of lattice
sites based on the value of the site's spin)~\cite{personal:parisi}.

Let us then divide the particles into mobile
($\x=\{\x_1,\ldots,\x_M\}=\{\r_1,\ldots,\r_M\}$) and frozen
($\y=\{\y_1,\ldots,\y_{N-M}\}=\{\r_{M+1},\ldots,\r_N\}$). Then the
point-to-set is
\begin{equation}
  C(0,\bar\Reg) = \langle \rho(0;\x) f(\{\y\}) \rangle_\Reg, \qquad 
  \mbox{with}\quad  f(\y)=\langle \rho(0;\x,\y) \rangle_{\Reg,\y},
  \label{eq:point-to-set-offlattice}
\end{equation}
where in our case $\rho(\r)$ is the coarse-grained density
\begin{equation}
  \rho(\r;\x)=\int_{v_\r} d^3s \sum_i^M \delta(\vec{s}-\x_i),
\end{equation}
($v_\r$ is a small volume at the centre of $\Reg$) and
$\langle\ldots\rangle_\Reg$ means an average with a
hard wall around $\Reg$, \textsl{i.e.\/}
\begin{eqnarray}
\langle\ldots\rangle_\Reg &=& \frac{1}{\Zpart}\int\!\!d^M\x d^{N-M}\y\,
\ldots \boltzm{\x}{\y}, \nonumber\\
\tilde H(\x,\y)&=&H(\x,\y) + \sum_i v_{\mathrm{in}}(\x_i) + \sum_i
v_{\mathrm{out}}(\y_i),
\end{eqnarray}
where $v_{\mathrm{in}}(\r)$ is a potential that is 0 inside $\Reg$ and
infinite outside (and conversely for $v_{\mathrm{out}}$).
Then we can write the expression corresponding
to~(\ref{eq:point-to-set-expanded}),
\begin{eqnarray}
  C(0,\bar\Reg) &=& \frac{1}{\Zpart} \int\!\! d\x d\y \boltzm{\x}{\y}
    \rho(0;\x) \frac{1}{\Zpart_\y} \int\!\! d\x' \boltzm{\x'}{\y}
    \rho(0;\x') \nonumber \\
   &=& \int\!\!d\y\, \frac{\Zpart_\y}{\Zpart} \frac{1}{\Zpart_\y}
   \int\!\!d\x \rho(0;\x) \boltzm{\x}{\y}
   \frac{1}{\Zpart_\y} \int\!\!d\x'  \rho(0;\x') \boltzm{\x'}{\y}
   \nonumber\\
   &=& \overline{ \left\langle \rho(0;\x) \right\rangle_\y^2 } .
   \label{eq:point-to-set-offlattice-expanded}
\end{eqnarray}

\section{Numerical results: statics}

\label{sec:numer-results:-stat}

\begin{figure}
  \includegraphics[width=\columnwidth]{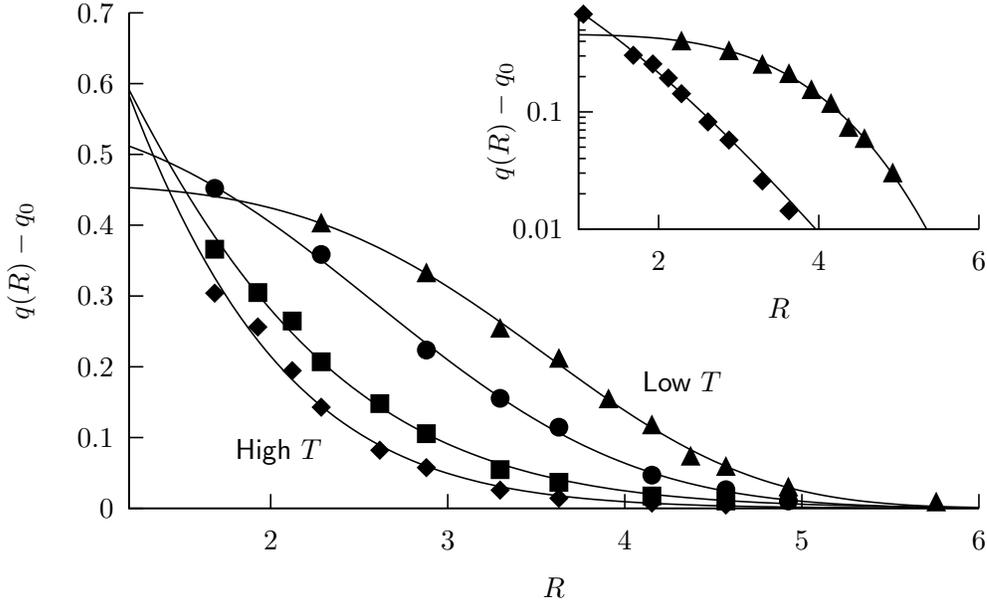}
  \caption{Point-to-set correlation (overlap) vs.\ cavity radius for
    the soft-sphere binary mixture at temperatures $T=0.482$
    (diamonds), 0.350 (squares), 0.246 (circles), and 0.203
    (triangles) \cite{self:nphys08}. Lines are fits to the compressed
    exponential form~(\ref{eq:qvsR-compressed}). Inset: Same data for
    the highest and lowest temperatures in log-linear scale, showing
    the non-exponential character of the overlap decay at low
    temperatures (an exponential shows as a straight line in this
    plot).}
  \label{fig:xivsr}
\end{figure}

An off-lattice glassformer, the soft-sphere binary mixture, has been
recently studied numerically under \acp{ABC}
\cite{self:prl07,self:nphys08}. In particular, the point-to-set
correlation~(\ref{eq:point-to-set-offlattice-expanded}) was computed
using Monte Carlo simulations in ref.~\cite{self:nphys08}. In this
work, \acp{ABC} allowed for the first time to detect a clearly growing
\emph{static} correlation length (figure~\ref{fig:xivsr}).

Another finding of ref.~\cite{self:nphys08} is that at low
temperatures the decay of the overlap (or point-to-set correlation) is
\emph{non-exponential} (figure~\ref{fig:xivsr}, inset), and can
be fitted by the empirical form (compressed exponential)
\begin{equation}
  \label{eq:qvsR-compressed}
  q(R)-q_0 = A\exp\left[-\left(R/\ptsl\right)^\zeta\right],
\end{equation}
where $q_0$ is the asymptotic value of the overlap.

The growth of $\ptsl$ as temperature is decreased (and correlation times
grow) is a result not unexpected (even though it took many years to
develop the tools to detect the correlation length) in the sense that
it corresponds to physical intuition. As such, at the qualitative
level it does not constitute a new constraint on theories of the
supercooled liquid state. Instead it is the details of the $\tau$
\textsl{vs.\ }$\ptsl$ relationship, as well as the decay of the overlap
with $R$ which are expected to provide valuable information to help
discriminate among competing theories.

While the question of which theory best describes de supercooled
liquid is far from settled, the numerical data on the decay of the
overlap with $R$ under \acp{ABC}  have been shown to be
compatible with (a slightly generalized version of) the \ac{RFOT}
\cite{mosaic:kirkpatrick89, mosaic:dzero05, review:lubchenko07}
According to \ac{RFOT}, whether or not a region of radius $R$
relaxes depends on the balance between the surface tension $Y$ that
develops when that region actually rerranges and the configurational
entropy $\Sigma$ unleashed by the rearrangement: if $Y>T\Sigma
R^{d-\theta}$ ($d$ is the dimension of the system and $\theta$ the
surface tension exponent) the surface cost is larger
than the entropic gain and the region does not rearrange.  On the
other hand, if $Y<T\Sigma R^{d-\theta}$ the entropic gain outweighs
the surface energy cost and the region has a thermodynamic advantage
to rearrange. The minimal rearranging size where entropy and surface
tension balance, $\xi = \left(Y/T\Sigma\right)^{1/(d-\theta)}$, is the
mosaic correlation length of RFOT. Introducing a distribution $P(Y)$
that allows for fluctuations of the surface tension, a decay of the
overlap of the form~(\ref{eq:qvsR-compressed}) is
obtained~\cite{self:nphys08}.

The existence of a surface energy cost required by \ac{RFOT} for the
situation when different states are in contact might seem in
contradiction with~(\ref{eq:one-time}), which in particular means that
the energy is completely unaffected by \acp{ABC}. For a given \ac{BC},
the cavity explores all possible configurations, many of which should
pay a substantial energy price. Why does not this price show up in the
average energy? The answer is that precisely because of this high
price, those configurations are seldom visited: this is the pinning
induced by the \acp{BC}, which produces a high overlap for small
cavities. When the cavity is large the central region can rearrange
independently of the borders and the cavity becomes ergodic; for these
sizes \ac{RFOT}'s states are no longer well-defined. In order to
detect surface tension, different \emph{frozen} configurations must be
put in contact~\cite{self:jstatmech09,mosaic:zarinelli10}.

\section{Numerical results: dynamics}

\label{sec:numer-results:-dynam}

The dynamics of a system under \acp{ABC} has been studied in
refs.~\cite{mosaic:jack05} and~\cite{self:arxiv10}. It is of interest
to establish not only the infinte time limit of the overlap (the
point-to-set value), but also how long it takes to reach the
asymptotic level. This is important in two respects, and again the
detailed answer is expected to provide constraints for the theories of
the liquid state.

First, one asks whether the point-to-set length $\ptsl$ is the
relevant dynamic length, \textsl{i.e.\ }whether its growth actually
tracks closely the growth of the correlation time $\tau$. The rigorous
bounds are rather broad: $\tau$ must grow at least linearly with
$\ptsl$ and no more than exponentially in
$\ptsl^d$~\cite{dynamics:montanari06} ($d$ is the space
dimension). The lengthscale directly linked to $\tau$ might be another
one. Another way to put the question is to ask whether $\ptsl$ is the
typical size of a \ac{CRR}: a \ac{CRR} is a region made of particles
that can only relax if they move
together~\cite{glassthermo:adam65}. Thus a region larger than
the size of \acp{CRR} will not relax slower than a \ac{CRR}, because
different regions can relax in parallel: this is thus the relevant
dynamic length.  Although \acp{CRR} \emph{cannot be smaller than
  $\ptsl$} (because, by definition, regions smaller than $\ptsl$
cannot fully relax), a region of this size could still relax
differently from the bulk.

The second question is whether the cavity correlation time approaches
the bulk time from above or from below.  Theories based on kinetic
constraints, where the slowdown is ascribed to the low density of
mobile regions (termed {\em excitations} or {\em defects})
~\cite{heterogeneities:merolle05,review:chandler10}, seem to imply that the
bulk equilibration time should be approached from above
(\textsl{i.e.\/} small systems relax more slowly).  On the other hand,
according to \ac{RFOT} the cavity equilibration time is expected to
have the opposite behaviour: when frozen amorphous boundary conditions
are present, small systems relax faster than large systems
\cite{review:biroli09}.

The (on-lattice) \ac{SPM} was studied in~\cite{mosaic:jack05}. This
model's dynamics is ruled by defects, and figure~8 of
ref.~\cite{mosaic:jack05} shows clearly that the bulk corerlation time
is reached from \emph{above,} while in figure~7 of the same work it
can be seen that the correlation time does not reach the bulk value
for the cavity size where the point-to-set correlation decays to 0
(\textsl{i.e.\ }$\ptsl$).

\begin{figure}
  \includegraphics[width=\columnwidth]{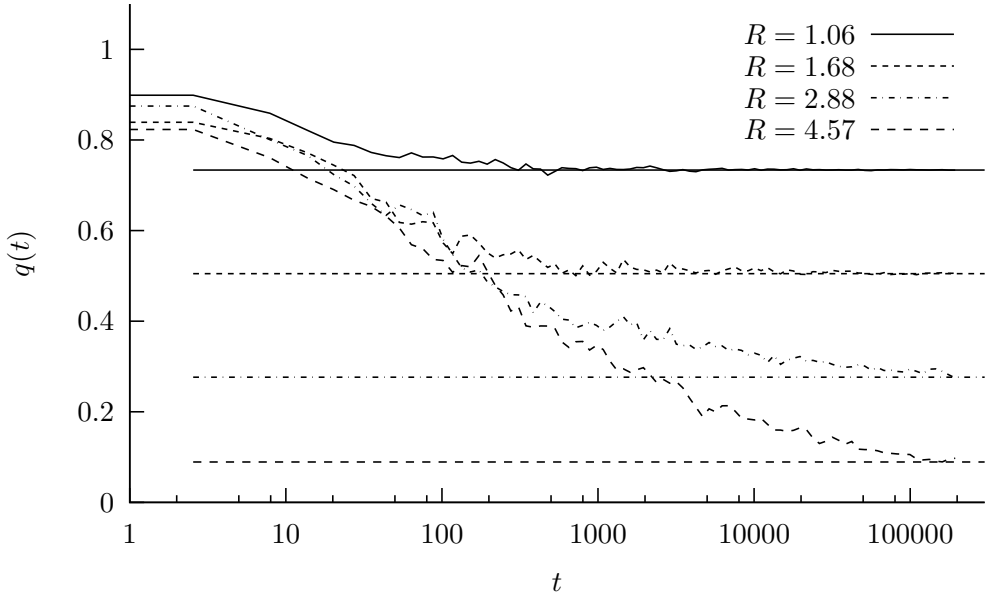}
  \caption{Time dependent overlap
    (see~(\ref{eq:autocorr-inf-time-limit})) as a function of time
    for four cavity sizes for the soft-sphere binary mixture at
    $T=0.350$ (see ref.~\cite{self:arxiv10}). As the point-to-set
    correlation (asymptotic value of $q(t)$) becomes lower, the time
    to reach the asymptotic value becomes larger.}
  \label{fig:qvst-nonconnected}
\end{figure}

\begin{figure}
  \includegraphics[width=\columnwidth]{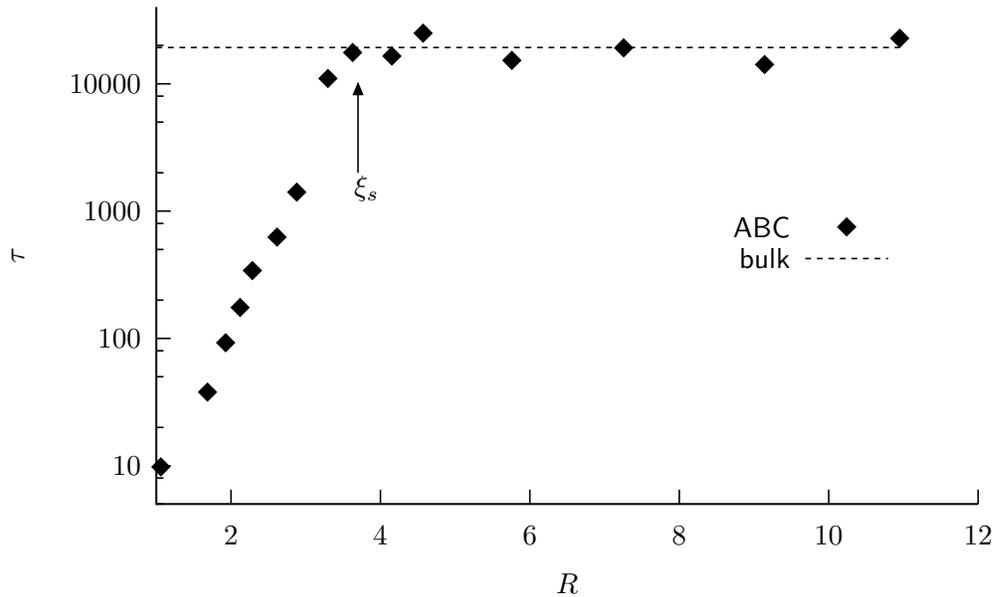}
  \caption{Cavity equilibration time vs.\ cavity size for the
    soft-sphere binary mixture at $T=0.203$~\cite{self:arxiv10}. The
    bulk value is reached when the cavity has the size of the static
    correlation length. Similar results were found for other
    temperatures~\cite{self:arxiv10}.}
  \label{fig:tauvsR}
\end{figure}

On the other hand, in the soft-sphere binary mixture considered
in~\cite{self:arxiv10} the behaviour is the opposite: the correlation
time grows with cavity size and approaches the bulk value from
\emph{below,} as is clear in figures~\ref{fig:qvst-nonconnected}
and~\ref{fig:tauvsR}. 

Figure~\ref{fig:tauvsR} also shows that the bulk relaxation time is
reached at the point-to-set length $\ptsl$, \textsl{i.e.\ }that
$\ptsl$ is the relevant dynamic length. Thus in~\cite{self:arxiv10} it
was concluded that $\ptsl$ can indeed be interpreted as the typical
size of \acp{CRR}.

\section{Conclusions}

\label{sec:conclusions}

In summary, numerical simulations with \aclp{ABC} provide a new tool
to study systems with amorphous order, such as supercooled
liquids. The results we have discussed here support the interpretation
of $\ptsl$ as the relevant dynamic length. This corresponds to the
characteristic length $\xi$ of the \acf{RFOT} of liquids. Numerical
results at low temperatures can be interpreted within an \ac{RFOT}
scenario which allows for surface tension fluctuations, while the
correlation time that grows with $R$ found in those studies is
somewhat harder to explain in the context of defect-based
theories.

However, the debate about which theory best describes supercooled
liquids is not settled. We expect that numerical and
theoretical exploration of situations with \acp{ABC} will provide
highly valuable information and insight into the statistical mechanics
of these systems, and contribute to the efforts to build a
satisfactory theory.

\ack

We thank G.~Biroli, J.-P.~Bouchaud, C.~Cammarota, S.~Franz,
G.~Gradenigo and G.~Parisi for discussions. The work of TSG was
supported in part by grants from ANPCyT, CONICET, and UNLP
(Argentina). PV has been partly supported through Research Contract
Nos. FIS2009-12648-C03-01,FIS2008-01323 (MICINN, Spain).

\section*{References}

\bibliographystyle{jstat}
\bibliography{statphys}

\end{document}